\documentclass[twocolumn,showpacs]{revtex4}
\usepackage{dcolumn}
\usepackage{epsfig,epsf}

\newcommand{\bee}{\begin{equation}}
\newcommand{\ee}{\end{equation}}
\newcommand{\bea}{\begin{eqnarray}}
\newcommand{\eea}{\end{eqnarray}}
\newcommand{\R}{\rm I\kern-.2emR}
\newcommand{\C}{\rm \kern.25em\vrule height1.4ex
depth-.12ex width.06em\kern-.31em C}
\newcommand{\N}{{\rm I\kern-.16em N}}
\newcommand{\Z}{{\rm Z\kern-.35em Z}}
                                                      
\begin{document}

\renewcommand{\topfraction}{0.9}
\renewcommand{\textfraction}{0.1}
                                                                
\title{Lattice artefacts and the running of the coupling constant.
}

\author{Adrian Patrascioiu}
\affiliation
{\it Physics Department, University of Arizona,
 Tucson, AZ 85721, U.S.A.}
\author{Erhard Seiler}
\affiliation{\it Max-Planck-Institut f\"ur Physik
(Werner-Heisenberg-Institut)
F\"ohringer Ring 6, 80805 Munich, Germany}
\date{\today}
\date{\today}

\begin{abstract}
We study the running of the L\"uscher-Weisz-Wolff (LWW) coupling constant
in the two dimensional $O(3)$ nonlinear $\sigma$ model. To investigate
the continuum limit we refine the lattice spacing from the $1\over 16$
value used by LWW up to $1\over 160$. We find much larger lattice 
artefacts than those estimated by LWW and that most likely the coupling 
constant runs slower than predicted by perturbation theory. A precise 
determination of the running in the continuum limit would require a 
controlled ansatz of extrapolation, which, we argue, is not presently 
available.
\end{abstract}
\maketitle
The hallmark of $QCD$ is its alleged asymptotic freedom (AF), that
property which expresses the fact that at shorter distances the
interactions between quarks and gluons are becoming weaker. This
fact however has been established only in perturbation theory (PT),
an approximation scheme without a mathematical basis and which,
moreover, has been shown to be plagued by ambiguities (expectation
values of variables of compact support depend upon the boundary
conditions (b.c.) used to reach the thermodynamic limit) \cite{super}.

It is therefore most important to establish whether AF is really a property
of $QCD$ in a nonperturbative framework. The first step in this 
direction was taken in 1991 by L\"uscher, Weisz and Wolff (LWW) \cite{lww},
who proposed a method to investigate the presnce of AF in the two dimensional
($2D$) $O(N)$ nonlinear $\sigma$ models, which, perturbatively, are
also AF for $N\geq 3$. As a coupling constant they proposed the
following renormalization group invariant:
\bee
          \tilde g^2(\beta,L)={2L\over (N-1)\xi(L)}
\ee
Here $\xi(L)$ is the correlation length of the $O(N)$ model in an infinite
strip of width $L$ with periodic b.c.. 
It is defined by the following double limit: consider a finite strip
of size $L\times T$ with periodic b.c. in the direction of size $L$ and
arbitrary b.c. in the direction of size $T$. Then
\bee
\xi(L)=-lim_{x\to\infty} lim_{T\to\infty} {x \over\ln(<s(0)\cdot s(x)>)}.
\ee
For the standard $O(N)$
action
\bee
          H_{i,j}=-s(i)\cdot s(j)
\ee
at inverse coupling constant $\beta$, if in eqn.(1) one expressed $L$ in
units of the thermodynamic correlation length $\xi_{\infty}(\beta)$,
then in the limit $L\to\infty$ and $\beta\to\beta_{crt}$, holding
$z=L/\xi_{\infty}$ fixed, one would obtain a unique function 
$\bar g^2(z)$ describing the running of $\bar g^2$ with the physical 
distance $z$.

As we pointed out in 1992 \cite{run}, the interpretation of
$\bar g^2$ as a coupling constant is somewhat misleading since it does 
not measure the strength of the interaction. Indeed in the
the massive continuum limit of a free field theory  
$\bar g^2(z)$ is a nontrivial function, running linearly
with $z$.  LWW argued in favour of their
choice by pointing out that in PT, to lowest order in the bare (lattice)
coupling constant $\bar g^2(L)\sim {1\over \beta}$. This argument is
also problematic since even if $\beta_{crt}=\infty$, to construct
the continuum limit one would have to let $L\to\infty$ and thus reach
a regime where PT in the bare coupling would clearly not be applicable.

Nevertheless $\bar g^2(L,\beta)$ is a renormalization group
invariant and if one would discover that for some $\beta<\infty$,
$\bar g^2(L)$ became independent of $L$ for large $L$, that would mean that at
that $\beta$ the model is critical, which, as we will explain below,
would rule out the existence of AF in the massive continuum limit.

In 1991, for the $O(3)$ model,
 LWW claimed to be able to establish the continuum running of 
$\bar g^2(L/\xi_{\infty}(\beta))$ up to physical distances as small
as 0.03304(4) and to verify that it approached the perturbative (AF)
prediction. To achieve this they employed a finite size scaling (FSS)
technique: one measures $\xi(L)$ at some $\beta$ and $L$, then leaving
$\beta$ unchanged, one doubles $L$ and measures $\xi(2L)$. One thus
obtains a scaling curve giving $\bar g^2(2L)/\bar g^2(L)$ versus
$\bar g^2(L)$. This {\it step scaling curve}, as LWW called it,
allows one to connect small physical distances to large ones 
($L/\xi_{\infty}(\beta)>7$), where the continuum limit of   
$\bar g^2(L/\xi_{\infty}(\beta))$ could be reached.

Similar FSS approaches were used in 1993 by Kim \cite{kim} and in 1994 by
Caracciolo et al \cite{car} to predict the value of $\xi(\beta)$ up to
$\xi\sim 10^5$, even though the largest lattices involved did not 
exceed $L=512$. However whereas these authors produced their
scaling curves simply by observing that the Monte Carlo (MC) data 
coming from different values of $\beta$ seemed to fall on the same
curve, which they took as their step scaling curve, the LWW paper
claimed to have really controlled lattice artefacts (the aproach to
the continuum). More precisely the problem is this: of course if one knew
the continuum value of the step scaling curve one could connect small
physical distances to large ones. But in an MC investigation, by
necessity, one can only gather data at finite cutoff ($1/\xi_{\infty}(\beta)$
or alternatively $1/L$). However the continuum limit requires letting
$\xi_{\infty}(\beta)\to\infty$ and $L\to\infty$. Therefore one must in
principle worry about extrapolating the results obtained for the 
step scaling function at finite cutoff
to the continuum limit.

This feat, which the other above quoted authors did not even attempt,
was achieved by LWW by assuming a Symanzik type of approach to the
continuum limit. Namely at fixed $\bar g^2$ they assumed that the
step scaling function approaches its continuum limit value as $1/L^2$
(strictly speaking, inspired by PT, the Symanzik fit involves 
$log(L)/L^2$, however for $6\leq L\leq 16$ the log can be approximated by
a constant).
They backed this assumption with their MC data. However the values of
$L$ they used ranged only from $6\leq L\leq 16$. The main point
of our paper is to show numerical evidence that if one goes to 
much larger $L$ values, the Symanzik fit does not work and an
entirely new picture for the running of $\bar g^2$ with the physical
distance emerges.

Before showing our data, we would like to elaborate on a subtlety
having to do with the numerical determination of the correlation
length $\xi(L)$. Namely while the definition in eq.(1) is
mathematically well defined, one must adopt a computational 
procedure for implementing it. LWW used an $L\times T$ lattice with
free b.c. in the $T$-direction, took $T=5L$ and assumed a pure
exponential decay for $L<x<T-L$ (see Appendix B of ref.\cite{lww}).
We used instead the following numerical estimate
of the correlation length $\xi$: let $P=(p,0)$,
$p={2n\pi\over T}$, $n=0,1,2,...,T-1$, $T=10L$. Then
\bee
  \xi={1\over 2\sin(\pi/T)}\sqrt{ G(0)/G(1)-1 }
\ee
where
\bee
        G(p)={1\over LT}\langle |\hat s(P)|^2\rangle;\ \
\hat s(P)=\sum_x e^{iPx} s(x)
\ee
 
\begin{figure}[htb]
\centerline{\epsfxsize=8.0cm\epsfbox{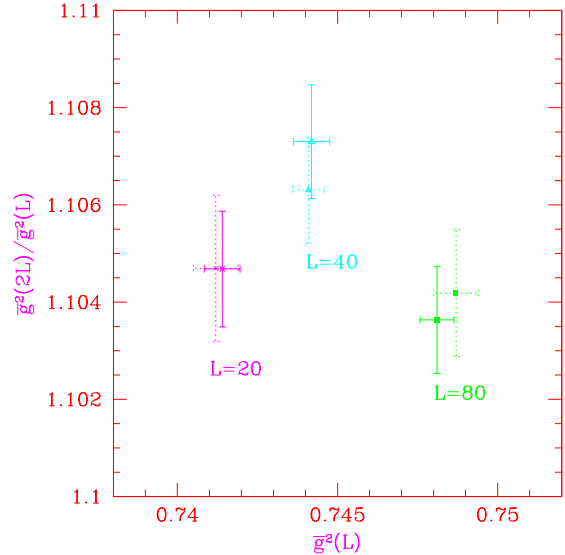}}
\caption{\it Step scaling function $\bar g^2(2L)/\bar g^2(L)$ versus
$\bar g^2(L)$ for $L=20,40,80$ computed with our method of estimating
$\xi(L)$ (solid lines) and with the LWW method (dotted line). }
\label{compare}
\end{figure}

It is not clear whether the LWW prescription or the one adopted by us
provides an estimate closer to the true exponential correlation 
length. For finite $T$ the LWW
procedure, employing free b.c.,
 is likely  to produce a value smaller than the true 
$\xi(L)$. In the procedure adopted by us, there are two effects of 
opposite sign, which could bring the result closer to the 
true exponential correlation length: \\
1. The periodic b.c. in the $T$-direction increases the order in
the system compared to an infinite strip.\\
2. Since our definition is sensitive to the multiparticle
states, for $T\to\infty$ it would produce a value smaller than the true
$\xi(L)$. This effect was studied by Campostrini et al
\cite{cp} using the high temperature expansion and
found to be less than 0.2\%.\\
Since we want to compare our data with those of LWW, we show in
Fig.\ref{compare} the resulting step scaling functions computed
with the two procedures at the same $\beta$ and $L$. As can be
seen, within numerical accuracy, at these values of $\bar g^2$ the 
two procedures produce similar results. 

\begin{figure}[ht]
\centerline{\epsfxsize=8.0cm\epsfbox{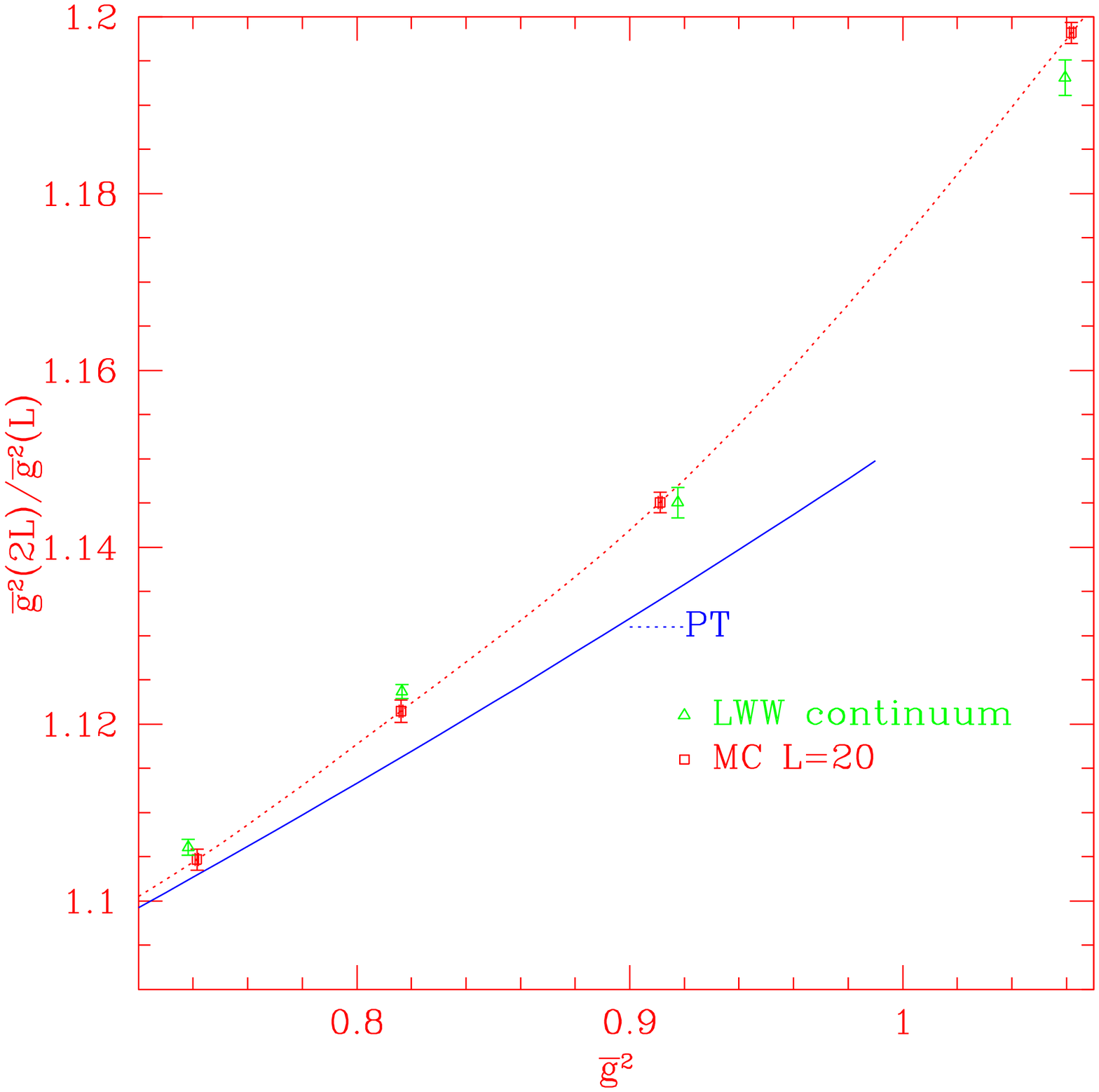}}
\caption{\it Step scaling function $\bar g^2(2L)/\bar g^2(L)$ versus
$\bar g^2(L)$ for $L=20$. Our MC data are connected by a spline.}
\label{step20}
\end{figure}

\begin{figure}[ht]
\centerline{\epsfxsize=8.0cm\epsfbox{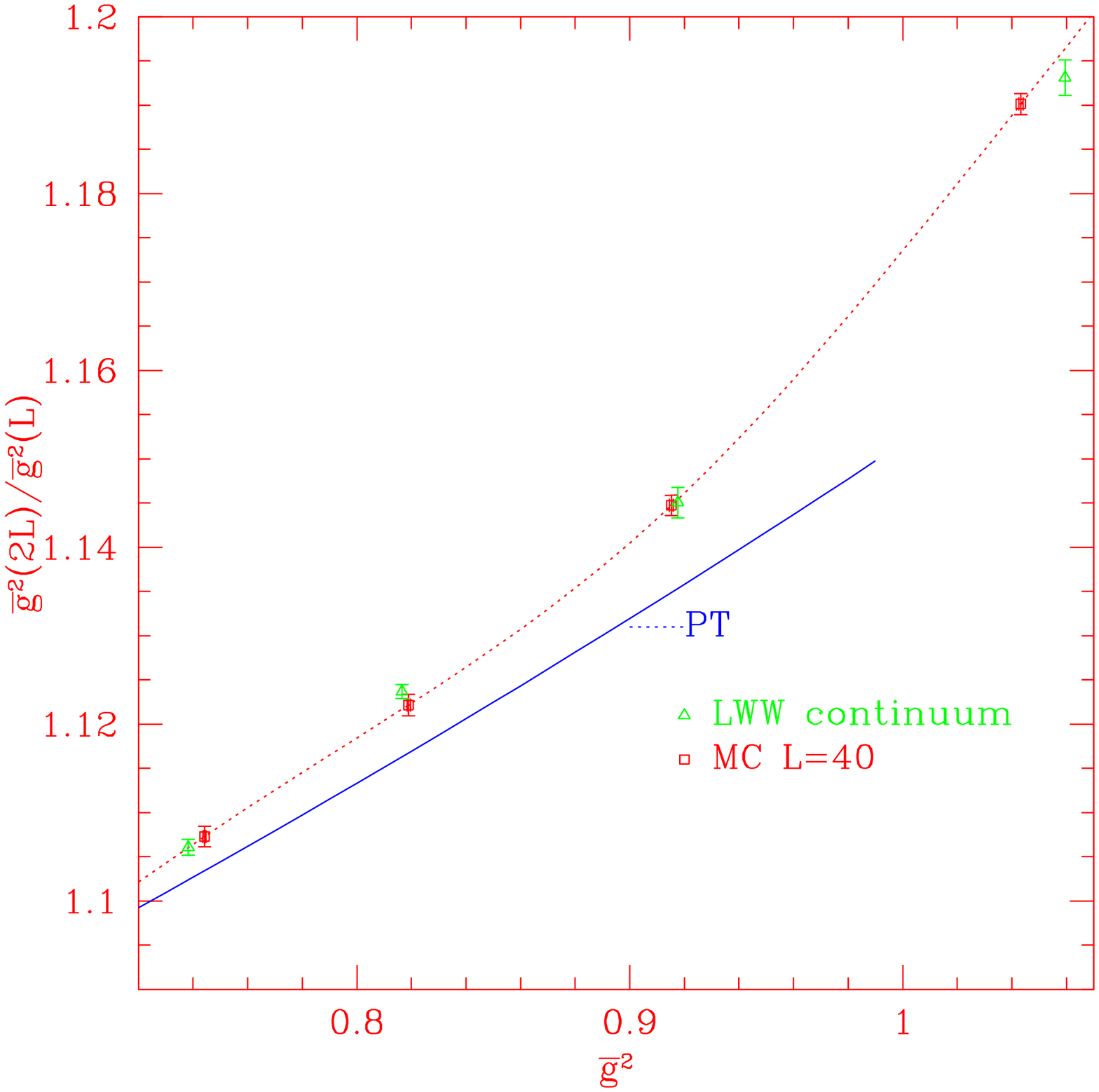}}
\caption{\it Step scaling function $\bar g^2(2L)/\bar g^2(L)$ versus
$\bar g^2(L)$ for $L=40$. Our MC data arew connected by a spline.}
\label{step40}
\end{figure}

\begin{figure}[ht]
\centerline{\epsfxsize=8.0cm\epsfbox{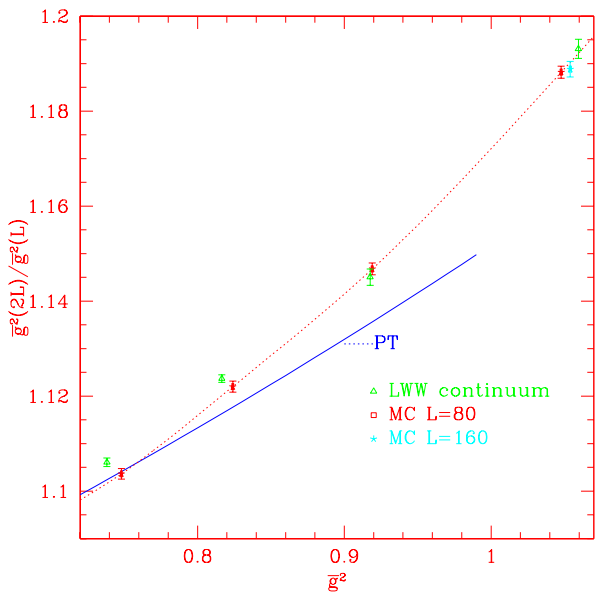}}
\caption{\it Step scaling function $\bar g^2(2L)/\bar g^2(L)$ versus
$\bar g^2(L)$ for $L=80$ and 160. Our MC data for $L=80$ are connected
by a spline.}
\label{step80160}
\end{figure}

Our results for the step scaling function are shown in Fig.\ref{step20},
Fig.\ref{step40} and 
Fig.\ref{step80160} for $L=20,40$ and $80$ respectively. The values were
obtained by measuring the correlation length at given $\beta$ first
on a lattice of size $L\times 10L$, then on $2L\times 20L$. 
The values of $\beta$
used (1.815, 1.94, 2.05, 2.16, 2.27, 2.38 and 2.49) were chosen
so that $\bar g^2$ took values in the range covered by LWW. 
The figures contain also two benchmarks taken from LWW \cite{lww}:\\
- The 3-loop PT curve. \\
- The LWW estimated continuum values.\\
Fig.\ref{step80160} contains also our value for the step
scaling function obtained by doubling $L$ from 160 to 320 at
$\bar g^2=1.05397(81)$. The error bars were estimated as follows:
for each value of $\beta$ and $L$ we started from a randomly chosen
configuration and ran the improved cluster algorithm
\cite{wolff} using 100,000 clusters for thermalization and 1,000,000
clusters for measurements. We then repeated this procedure a minimum of
157 times, except for the value at $L=320$ which contains only 74 runs.
We computed the average value over these samples of $G(0)$ and $G(1)$ and
from them $\xi(L)$. Since $\xi(L)$ is a nonlinear function of
$G(0)$ and $G(1)$, we estimated the error for $\xi(L)$ and
$\bar g^2(L)$ from our sample of independent values by the jack-knife
method. The data for $\tilde g^2(L)$ that were used in the figures
are given in Tab.1.

Our data do not agree with the LWW prediction \cite{lww} for the step
scaling function. The latter were obtained by extrapolating the MC
values obtained by taking $L$ between $L=6$ and $L=16$ via a second
order polynomial in ${1\over L^2}$. We did not display the LWW values
for $6\leq L \leq 16$
because of lack of space. However when combined with our values, they
reveal a rather complicated approach to the continuum. This is not
an unexpected fact. Indeed if $L$ is sufficiently small at given
$\beta$, the (asymmetric) lattice is strongly ordered in the transverse 
(shorter) direction. On the other hand, as we emphasized several 
years ago \cite{cont}, since in the continuum limit one must let
$L\to\infty$ and the Mermin-Wagner theorem guarantees the
restoration of the $O(N)$ symmetry in that limit for any finite
$\beta$, clearly for $L$ sufficiently
large this `perturbative' regime, of spins highly ordered in the
transverse direction, cannot persist. 

In fact we showed in \cite{cont} that
even if $\beta_{crt}=\infty$, as predicted by PT, the spins
would cease to be highly ordered in the transverse direction for 
$L$ sufficiently large. Indeed bare PT itself provides a clue as to
the distance over which the spins remain well ordered since to
lowest order one has
\bee
        \langle s(0)\cdot s(x)\rangle=1-{N-1\over \beta}D(x)
\ee
and to a good approximation $D(x)={1\over 4}+{1\over 2\pi}\ln(x)$.
Thus bare PT suggests that spins are well ordered over distances
$O(\exp([2\pi\beta/(N-1)])$. On the other hand the AF formula
predicts $\xi=O(\exp[2\pi\beta/(N-2)])$. Thus at fixed physical
distance (${L\over \xi(L)}$), in taking the continuum limit one would
surely leave the regime in which PT in the bare coupling is applicable.

Returning now to the pattern of lattice artefacts, initially, 
if $\beta$ is large enough, at small enough $L$, they should
follow the Symanzik pattern used by LWW because the system is
essentially in a PT regime. This regime has nothing to do with
the true approach to the continuum, which occurs only when $L$ is
sufficiently large (for given $\beta$) so that the $O(N)$ symmetry
becomes approximately true. How the true continuum limit is approached
can be model dependent. For instance in the Ising model there
are good reasons to expect a $1/L^2$ leading behavior \cite{bar}. On the
other hand in the $O(2)$ model there are both theoretical \cite{kos}
and numerical reasons \cite{hasenb} to expect a ${1/\ln(L)}$
approach. 

It is reasonable to expect that the $O(3)$ model, enjoying a continuous
symmetry, behaves as the $O(2)$, not as the Ising model. We have
attempted a ${1\over ln(L)}$ fit to our data but could not obtain a
reliable continuum value. (We encountered a similar situation in
investigating the renormalized coupling $g_R$ in the $O(3)$ model
\cite{col2}. The case of the $O(2)$ model is simpler to handle because
the theory predicts both the continuum value and the leading correction
\cite{hasenb}.) Even though we do not have a prediction for the
continuum step scaling function, our results do not corroborate
the original prediction of LWW and suggest
that most likely the nonperturbative running of $\bar g^2$ is slower
than predicted by PT. This situation is consistent with, though in no way
proving, the existence of a transition to a massless phase at
finite $\beta_{crt}$, as argued by us recently \cite{abs}. In that
paper we also proved rigorously that for the standard action, the massive
continuum limit cannot be AF if $\beta_{crt}<\infty$. The result
follows from a Ward identity and the reflection positivity of
the standard action.

Finally, regarding the running of $\alpha_s(Q)$ in $QCD_4$, all we
can say is that the Symanzik type of fit for the approach to the continuum
has no justification there either. Indeed, that fit is inspired by PT.
If in fact lattice $QCD_4$ does undergo a deconfining zero temperature
transition at nonzero (bare) coupling, so that the running of
$\alpha_s(Q)$ does not follow PT, there is no reason to expect the 
lattice artefacts to follow the Symanzik ansatz. 
Therefore it would be very useful if the lattice community employed its   
resources to establish first the true cutoff effects and the true
running of $\alpha_s$ in the pure Yang-Mills theory by going to larger
$L$, before attempting to handle dynamical fermions; the latter
unavoidably can only be done on minuscule lattices, and using the 
Symanzik fit to extrapolate to the continuum can be misleading, as we 
have found. As we said many years ago \cite{exp},
we expect that in the four dimensional Yang-Mills theory as well
as in $QCD_4$ there exists a nontrivial fixed point and
that $\alpha_s(Q)$ runs slower than predicted by PT, with the
effect becoming pronounced by 1 TeV or less.

Acknowledgement: We benefitted from numerous discussions with
Peter Weisz regarding the LWW paper. AP is grateful 
to the Werner-Heisenberg-Institut for its
hospitality.

\begin{widetext}
\vglue5mm
\begin{center}
{\bf Tab.1:}
{\it Monte Carlo data for $\tilde g^2(\beta,L)$}\\
\begin{tabular}[t]{l||r|r|r|r|r|r}
$\beta$ & 1.815 & 1.94 & 2.05 & 2.16 & 2.27 & 2.38 \\
\hline
\hline
$L=20$&1.06185(78)&.91115(67)&.81616(70) &.74170(56)&  &   \\
\hline
$L=40$&1.27230(85)&1.04334(76)&.91530(68)&.81901(63)&.74419(57)&  \\
\hline
$L=80$&  &1.24174(85) &1.04778(72)&.91907(70)&.82404(60)&.74811(53)\\
\hline
$L=160$&  &  &1.24497(100)&1.05398(81)&.92457(69)&.82564(58) \\
\hline
$L=320$&  &  &   &1.2530(15)&  &  \\
\end{tabular}
\vglue2mm
\end{center}
\end{widetext}


\begin{thebibliography}{99}
%
\bibitem{super} A. Patrascioiu and E. Seiler, {\sl Phys.Rev.Lett.} {\bf 74}
(1995) 1924.
%
\bibitem{lww} M. L\"uscher, P. Weisz and U. Wolff, {\sl Nucl.Phys.} {\bf B359}
 (1991) 221.
%
\bibitem{run} A. Patrascioiu and E. Seiler, {\sl Nuovo Cim.} {\bf 107A}
(1994) 765.
%
\bibitem{kim} J.-K. Kim, {\sl Phys.Rev.Lett.} {\bf 70}
(1993) 1735.
%
\bibitem{car} S. Caracciolo, R. G. Edwards, A. Pelissetto and A. Sokal,
{\sl Phys.Rev.Lett.} {\bf 75} (1995) 1891.
%
\bibitem{cp} M. Campostrini, A. Pelissetto, P. Rossi and E. Vicari,,
{\sl Phys.Lett.} {\bf B402} (1997) 141.
%
\bibitem{wolff} U. Wolff, {\sl Nucl.Phys.} {\bf B334} (1990) 581.
%
\bibitem{cont} A. Patrascioiu and E. Seiler , {\sl J. Stat. Phys.} {\bf 89}
(1997) 947.
%
\bibitem{bar} M. N. Barber, in {\sl Phase Transitions and Critical
Phenomena} {\bf vol.8} ed. C. Domb and J. L. Lebowitz (1983).
%
\bibitem{kos} J. M. Kosterlitz, {\sl J.Phys.} {\bf C7} (1974) 1046.
%
\bibitem{hasenb} M. Hasenbusch, {\sl An Improved Estimator for the
Correlation Function of 2D Nonlinear Sigma Models} hep-lat/9408019 (1994).
%
\bibitem{col2} J.~Balog, M. Niedermaier, F. Niedermayer, A. Patrascioiu,
E. Seiler and P. Weisz, {\sl Nucl. Phys.} {\bf B576} (2000) 517.
%
\bibitem{abs} A. Patrascioiu and E. Seiler, {\sl Absence of
asymptotic freedom in nonabelian models} hep-th/0002153 (2000).
%
%
\bibitem{exp} A. Patrascioiu and E. Seiler, in {\sl Rencontres de Physique
de la Vall\'ee d'Aoste} ed. M.Greco, Editions Fronti\`eres (1992) p.125.
%
\end{thebibliography}
\end{document}